\documentclass{article}

\usepackage{amsmath, amsthm, amssymb, subfigure, wrapfig, verbatim}
\usepackage[T1]{fontenc}
\usepackage{color}
\usepackage{graphicx}

\newtheorem{theorem}{Theorem}
\newtheorem{lemma}{Lemma}
\newtheorem{defini}{Definition}

\newtheorem{cor}{Corollary}

\title{Detecting all regular polygons in a point set}

\author{Greg Aloupis\thanks{aloupis.greg@gmail.com}~~\footnotemark[6]
\and Jean Cardinal\thanks{jcardin@ulb.ac.be}~~\footnotemark[6]
\and S\'ebastien Collette\thanks{Charg\'{e} de Recherches F.R.S.-F.N.R.S.,
sebastien.collette@ulb.ac.be}~~\footnotemark[6]
\and John Iacono\thanks{Department of Computer  Science and Engineering,
Polytechnic Institute of New York University, 5 MetroTech Center, Brooklyn NY 11201. Research
supported in part by
NSF grants CCF-0430849 and OISE-0334653. http://john.poly.edu.}
\and Stefan Langerman\thanks{Ma\^{i}tre de Recherches F.R.S.-F.N.R.S.,
stefan.langerman@ulb.ac.be}~~\footnote{Dept. d'Informatique, Universit\'e Libre de Bruxelles, CP212,
Boulevard du Triomphe, 1050 Bruxelles, Belgium.}
}

\index{Aloupis, Greg}
\index{Cardinal, Jean}
\index{Collette, Sebastien}
\index{Iacono, John}
\index{Langerman, Stefan}

\begin{document}
\maketitle

\begin{abstract}
In this paper, we analyze the time complexity of finding regular polygons
in a set of $n$ points. We combine two different approaches to find
regular polygons, depending on their number of edges.  
Our result depends on the parameter $\alpha$, which has been used to bound
the maximum number of isosceles triangles that can be formed by $n$ points. 
This bound has been expressed as $O(n^{2{+}2\alpha{+}\epsilon})$, and the current best value
for alpha is $0.068$.

Our algorithm finds polygons with 
$O(n^{\alpha})$ edges  by sweeping a line through the set of
points, while  larger polygons are found by random sampling. We can
find all regular polygons with high probability in $O(n^{2{+}\alpha{+}\epsilon})$ expected time for
every positive $\epsilon$.  This compares well to the
$O(n^{2{+}2\alpha+\epsilon})$ deterministic algorithm of Bra{\ss}~\cite{brass}.
\end{abstract}

\section{Introduction}

The focus of this study is on the detection of regular structure in point sets.
Our motivation comes from observations that have been published concerning
extraordinary symmetries in the placements of ancient towns, temples and other
important locations. 
For instance, the oracle of Delphi has been measured to
be the apex of isosceles triangles with at least seven pairs of ancient Greek
cities\footnote{For an informal, illustrative and detailed account, see\newline \tt{http://www.geocities.com/sfetel/en/geometry.htm}.}.  The same is true for the oracle at Dodoni, while the small island
of Delos is the apex of at least thirteen isosceles triangles.  All three of the
central locations were considered to be among the most important of places, and
in fact Delphi was considered to be the navel of the world.  In general, one may
find seemingly countless cases of collinearity, reflective symmetry,
partial $n$-gons and networks of 
isosceles triangles when looking at the graph of cities in the ancient world, 
from the British Isles to the Middle East.  

We will not concern ourselves further questioning whether such structures were
carefully constructed or instead an expected result on large complete geometric
graphs.  However, the topic generates other interesting questions.
If one chooses a particular location as a temple, it is not 
difficult to construct cities (at least on paper)
so that the temple becomes the center of several symmetries.
What about the opposite?  Given a set of existing cities, where should one decide
to place a temple?  Or, to ask differently, where should one look for a hidden
temple?
\begin{figure}[t!]
\center\includegraphics[scale=.5]{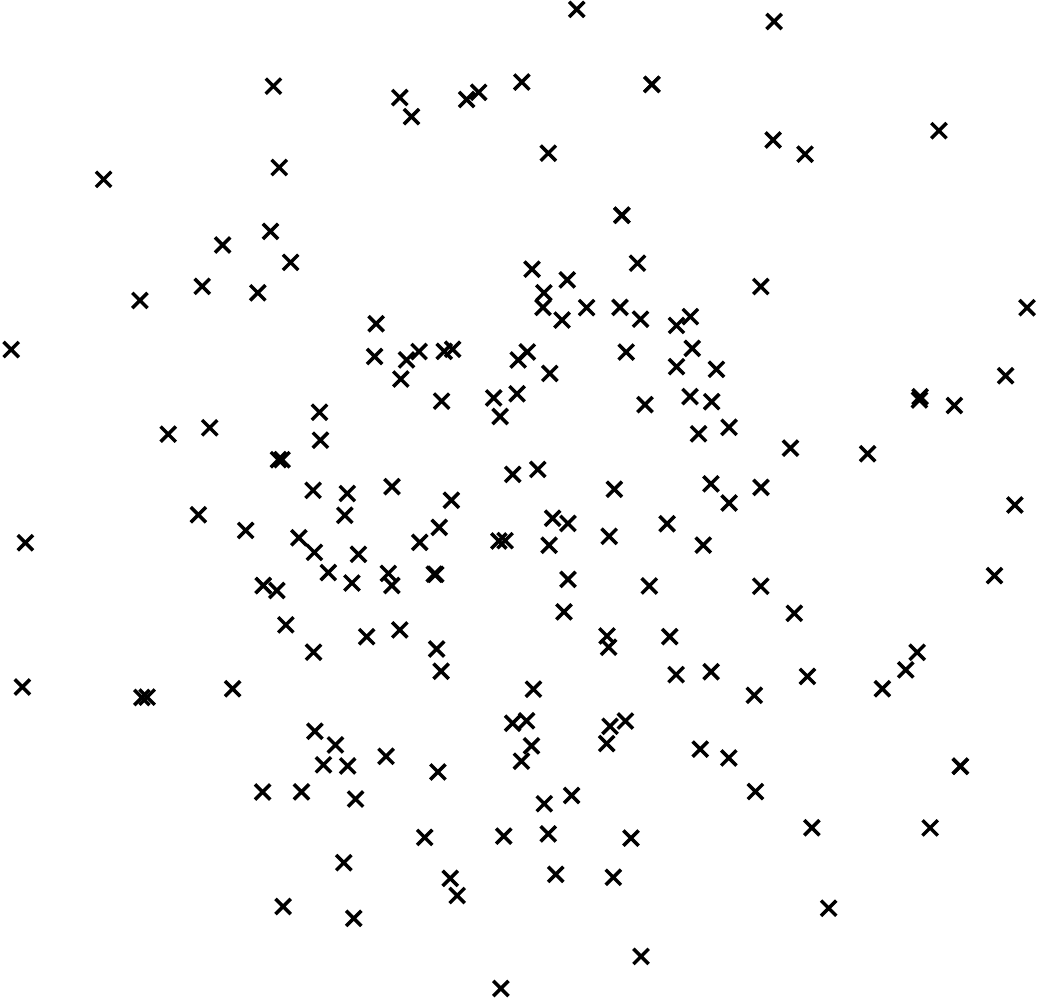}
\caption{Can you find all regular $k$-gons in this figure? Solution
is in Figure~\ref{jf2}.}
\label{jf1}
\end{figure}

\section{Related Work and Statement of New Result}
\label{previous}

Given a set of $n$ points, we wish to find the maximum subset which satisfies
a specific symmetry or structure.

The algorithm by Bra{\ss} \cite{brass}, for 
finding maximum symmetric subsets in point sets, is capable of
handling reflective lines, translations, rotational symmetries and repeated sets.
The time complexity is $O(n^{2.136{+}\epsilon})$ for every positive $\epsilon$.

The bound of Bra{\ss} depends on the maximum number of
isosceles triangles formed by a set of points in the plane. Pach and
Agarwal \cite{pachagarwal} bound this value by $O(n^{2{+}1/3})$. 
This was improved by Pach and Tardos
\cite{pach} to $O(n^{\frac{11e{-}3}{5e{-}1}{+}\epsilon}) \simeq O(n^{2.136{+}\epsilon})$.

Bra{\ss} noted that another result of his algorithm
 was to find all regular polygons contained in the set.
We improve the time complexity for detecting regular polygons in point
sets, although unlike the algorithm by Bra{\ss}, our algorithm is randomized.

Our  bound is $O(n^{2{+}\alpha+\epsilon})$, where $\alpha{\simeq}0.068$.
Notice that $\alpha$, the fractional component in the exponent of $n$, is half of the equivalent
component in~\cite{brass}. This is no coincidence.
Our algorithm is designed to reduce this fraction by a factor of 2.
Thus, any improvement of the result of Pach and Tardos in~\cite{pach} will be directly
reflected in our algorithm.

\section{Model of Computation}
\label{model}
We assume that all coordinates and other values are stored in a format
that allows constant time equality testing and hashing. As hashing is
only used to speed up one dimensional searches, it can be substituted
with a comparison-based structure. This increases the computational complexity by a 
logarithmic factor, which
is absorbed into the $\epsilon$. Furthermore, as exact computation
methods are typically not used, comparison based structures can be
used to substitute equality tests with proximity tests for
suitably small proximity values.  This will compensate for any small discrepancies in
the computation.

Note that we frequently use the variable $\epsilon$ inside asymptotic notation. Such claims hold for any constant $\epsilon >0$, and the asymptotic notation may hide constants that depend on $\epsilon$. Thus, for example, $f(n)=O(n^\epsilon \log n)$ if and only if $f(n)= O(n^\epsilon)$.

\begin{figure}[t!]
\center\includegraphics[scale=.5]{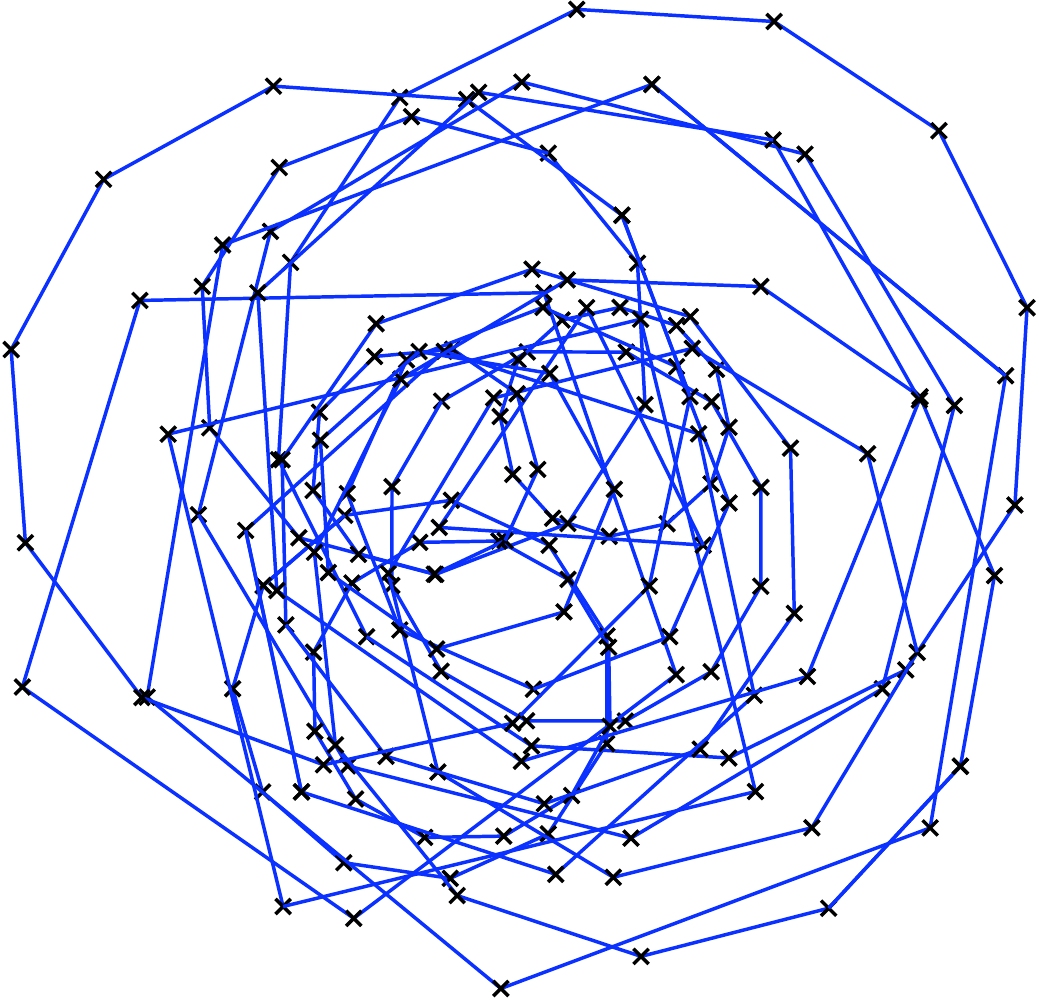}
\caption{All regular $k$-gons in the point set illustrated
in Figure~\ref{jf1}.}
\label{jf2}
\end{figure}

\section{Results}
We state our main theorem immediately.
\begin{theorem}
There is an algorithm which, with high probability, finds all  polygons in a set of $n$ points,
in expected time $O(n^{2{+}\alpha+\epsilon})$.
\end{theorem}
We obtain our result by considering small and large $k$-gons separately, obtaining the same bound in both cases.   Thus we have two separate proofs given in this section.
For $k\leq n^\alpha$, we give a sweep-line algorithm in section~\ref{subsec:small} (Lemma~\ref{l1}).
Larger polygons are found by the method given in section~\ref{subsec:big} (Theorem~\ref{big-gons}).
Let regular polygons of size greater than $n^\alpha$ be denoted as ${\geq}n^\alpha$-gons.
Smaller polygons are denoted as  ${\leq}n^\alpha$-gons.

\subsection{Small $k$-gons}
\label{subsec:small}
\begin{lemma}
\label{l1}
For any $\alpha$, we can 
find all ~${\leq} n^{\alpha}$-gons formed by $n$ points, in $O(n^{2{+}\alpha}\log n)$ time. 
\end{lemma}

\begin{proof}
Let $S$ be the given set of $n$ points.
We compute all line segments defined by pairs of points in $S$, and we
view this as an embedded graph. 
For each vertex, we construct a hash table 
containing incident edges, stored by key value and length.

Let $\phi_i = \pi-\frac{2 \pi}{i}$, and let $\Phi= \{ \phi_3, \phi_4, \ldots
\phi_{n^{\alpha}} \}$ be the set of all
angles formed by three adjacent vertices in a regular ${\leq}n^{\alpha}$-gon. 
Let $e{=}v_1v_2$ be an edge in the embedded graph.
We will show how to determine if any neighboring edge of $e$ might be
in the same ${\leq} n^{\alpha}$-gon as $e$, in  $O(n^{\alpha})$ time.
To do this, we use $\Phi$ and the hash tables of $v_1$ and $v_2$. 

The algorithm uses a simple left-to-right line sweep (see Figure~\ref{smallgons}). As we sweep, 
we propagate two types of messages along edges: ``possible
$k$-gon above/below''.
Edges may carry several messages at the same time.
 %
During the sweep, we stop at each vertex $v$ and 
process the following types of events:

\begin{itemize}
\item {Origination Event.} This occurs when $v$ might be the leftmost
vertex of a $k$-gon. If two edges of the same length are to the right of
$v$ and are at angle $\phi_{k}$, then we give a ``possible $k$-gon
below'' signal to the upper edge and a ``possible $k$-gon above" signal to the
lower edge. 

\item {Propagation Event.}
This event confirms or rejects a scenario in which $v$ is a possible vertex of a
$k$-gon (not the leftmost or rightmost).
The event is triggered when a ``possible $k$-gon above/below'' signal is
received from an edge $e$ incident and to the left of $v$.
If there is a right-facing edge $r$, of the
same length and forming an angle $\phi_k$ with $e$, 
the signal is propagated to $r$.  Naturally, the
orientation of the angle must match the type of signal (above or below). 
If there is no such right-facing edge, the
signal (and candidate $k$-gon) is discarded.

\item {Termination Event.} This event detects the rightmost
vertex of a $k$-gon. This happens if $v$ receives a
``possible $k$-gon above'' and a ``possible $k$-gon'' below signal
from two left-facing edges of the same length, angle $\phi_k$, and
in the proper relative orientation.
We output the description of the $k$-gon (the center, rotation, and value $k$ can
be easily determined from the information at hand), and we do not propagate
the two incoming signals.
\end{itemize}

\begin{figure}[h!]
\center
\includegraphics[scale=.5]{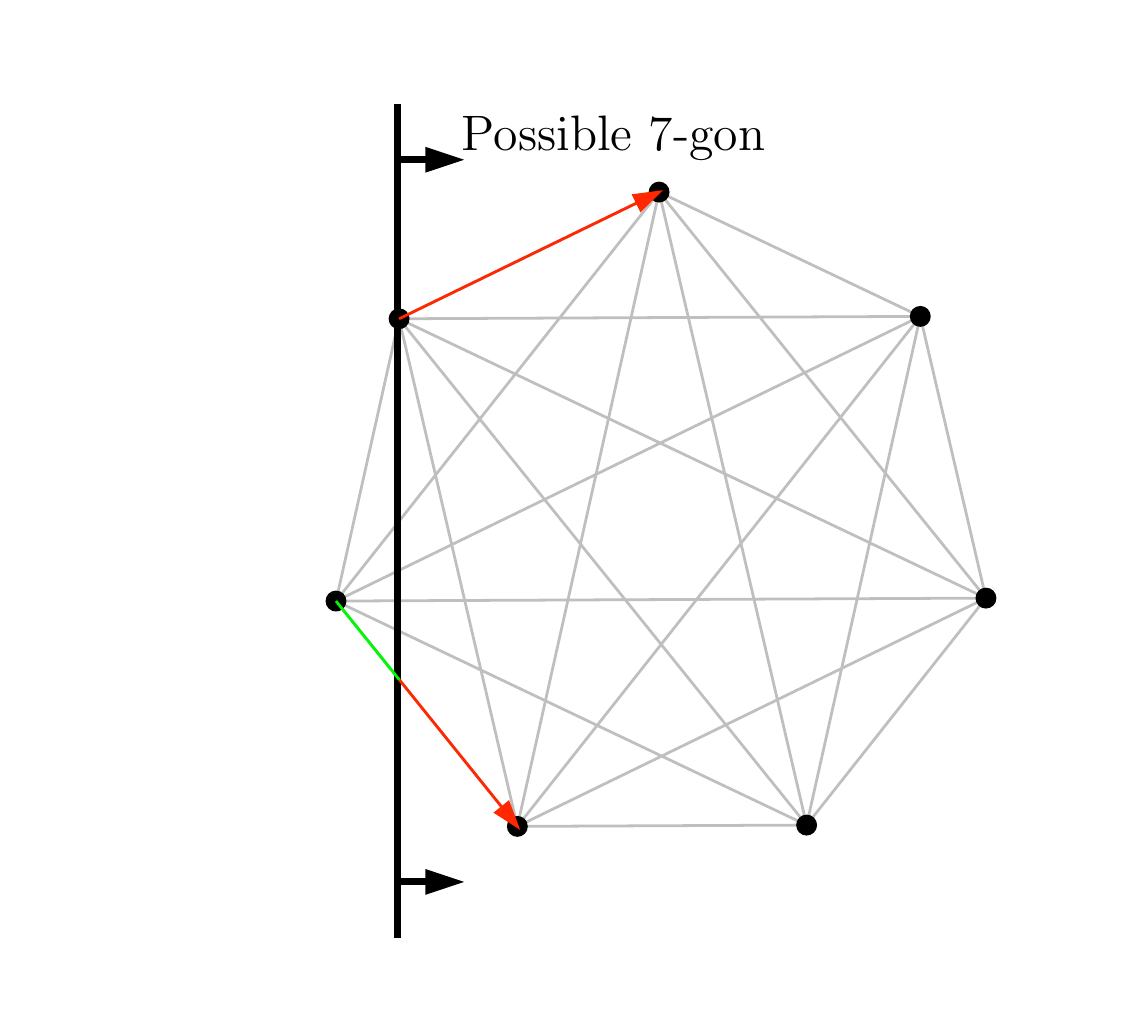}
\includegraphics[scale=.5]{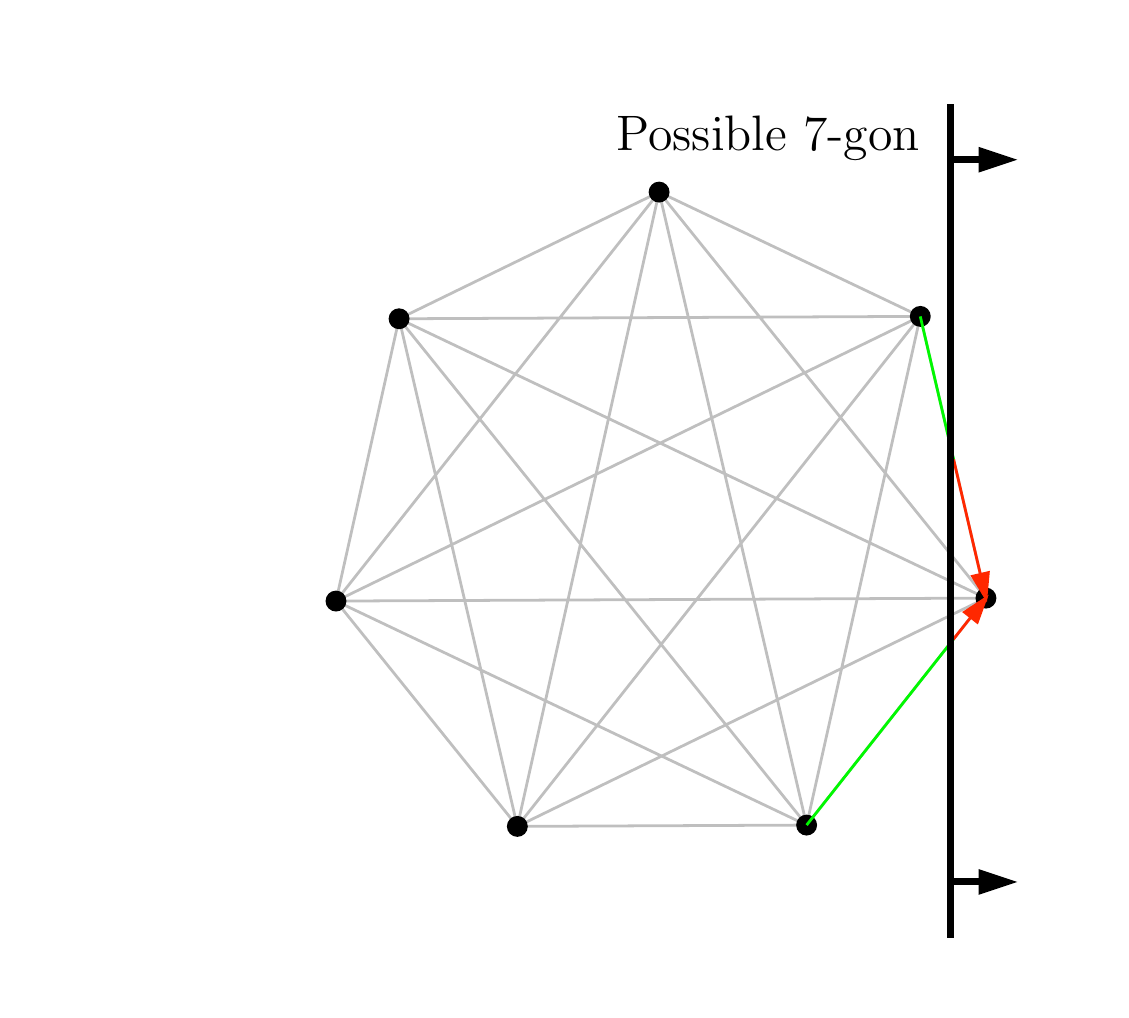}
\caption{Handling small $k$-gons.  Left: propagation;  Right: termination.}
\label{smallgons}
\end{figure}

The sweep exhaustively scans every possible ${\leq}n^\alpha$-gon.
If such a polygon exists, the origination event will identify its leftmost vertex.
Propagation will verify that the correct angles and edge lengths exist throughout
the polygon, and termination will match the top and bottom chains of the polygon.

 It takes $O(n^2 \log n)$ time to perform a line
sweep. (The astute reader will realize that a topological sweep~\cite{sweep} could be
used instead, at a cost of $O(n^2)$, but as all logarithmic factors get absorbed in 
the $\epsilon$, this is not critical). 
There are at most $2 n ^{\alpha}$ possible signal types, two for each polygon size.
If all signal types were to appear on each of the $O(n^2)$ edges, the total number of signals 
 over the entire sweep would be $O(n^{2{+}\alpha})$.
Propagation and termination events are done with table lookup and take constant time for each event. Generating all
origination events takes $O(n^{\alpha})$ table lookups for each incident edge to the right,
as all angles in $\Phi$ are searched. Thus, the total runtime is $O(n^{2{+}\alpha})$.
\end{proof}

\subsection{Large $k$-gons}
\label{subsec:big}
In this section we show how to identify all ${\geq}n^\alpha$-gons in $S$ with high probability.

Each such polygon will be quickly identified by its own {\em special} triangle.  In fact we show that each polygon is associated with several triangles that could uniquely identify it.   We sample triples of points in such a way that, with high probability, at least one special triangle will be chosen for each existing
$k$-gon.  Our algorithm can be summarized by the following:
\begin{itemize}
\item
Randomly select ``enough"
isosceles triangles from the point set.
\item
For each isosceles triangle, determine if it is such that it could uniquely identify a $k$-gon (i.e. if it is special).
\item
For every special triangle,  test if the vertices of its unique matching $k$-gon exist.
\end{itemize}

\noindent In the remainder of this section, we describe how to select random isosceles triangles 
(see~\ref{random}),  we describe special triangles and their properties (see~\ref{special}), and
we explain how to test for all $k$-gons given a sufficient set of special triangles (\ref{testing}).
Finally, we piece these elements together and describe the entire process that gives our result
(\ref{alltogether}).


\subsubsection{Random selection of isosceles triangles}
\label{random}
We start by explaining how to select a random apex of an isosceles triangle from the set of all apices of isosceles
triangles formed by a point set.   Note that we wish to do this without explicitly constructing
all isosceles triangles.
\begin{lemma}
\label{lem:random}
Let $S$ be a set of $n$ points. Three ordered points $p,q,r \in S$ form an \emph{isosceles
triple} if $|\overline{pq}| = |\overline{pr}|$.
With $O(n^2)$ preprocessing, an isosceles triple can be selected
uniformly at random from the set of all isosceles triples, in constant time.
\end{lemma}
\begin{proof}
Let $G$ be the complete geometric graph on $S$.
Let $e(p,\ell)$ be the set of segments of length $\ell$, incident to  
$p$:  $e(p,\ell)=\{ \overline{pq}{\in}G~|~|\overline{pq}|{=}\ell \}$.
 Let $B = \{ e(p,\ell)~|~|e(p,\ell)|{\geq}2 \}$. Let $I$ be the set of all isosceles triples.
Note that $|I|=\sum_{e(p,\ell) \in B} 2\binom{|e(p,\ell)|}{2}$.\\

Using bucketing, we can compute $e(p,\ell)$ (and $|e(p,\ell)|$) for all $p$
and $\ell$ in $O(n^2)$ time.
This can be done as follows. For each $\{p,q\}{\subseteq}S,~p{\not=}q$,
add the edge $\overline{pq}$ to the bucket with label
$e(p,|\overline{pq}|)$, creating the bucket if it does not exist.
The set $E$, of at most ${n \choose 2}$ buckets, is obtained by deleting all buckets with only one edge. Then, for each bucket $b \in E$ assign a
weight of $w(b)=\frac{{|b| \choose 2}^2}{\sum_{c \in B} {|c| \choose
2}^2}$. Note that $\sum_{b \in B} w(b)=1$. \\

Now, by selecting a random bucket $b(p,\ell)\in B$, as well as two
different random edges $\overline{pq}, \overline{pr} \in b(p,\ell)$, we obtain a random
 isosceles triple $(p,q,r)$. This can be done with three
random choices in time $O(1)$. The two edges $\overline{pq}$ and
$\overline{pr}$ both have length $\ell$ by virtue of being in
$e(p,\ell)$. Thus $(p,q,r)$ is guaranteed to be an isosceles triple.

Given an isosceles triple $(p,q,r)$, $\overline{pq}$ and
$\overline{pr}$ will appear in exactly one bucket $b(p,\overline{pq}=
\overline{pr})$ and the probability of picking this triple will be

$$\overbrace{\frac{{b \choose 2}^2}{\sum_{c \in B} {c \choose
2}^2}}^{\text{Pick a bucket}} 
~~~\cdot 
\overbrace{\frac{1}{ 2{b \choose
2}^2 }}^{\text{Pick two edges}} = 
~\frac{1}{|I|}$$

 \noindent Since  $|I|$ is the
number of isosceles triples this method chooses each triple
uniformly at random.  Our claim follows, since every isosceles apex corresponds to exactly
two ordered triples.
\end{proof}

\subsubsection{Special triangles:  characterization and properties}
\label{special}
\begin{figure}[h!]
\center
\includegraphics[scale=.49]{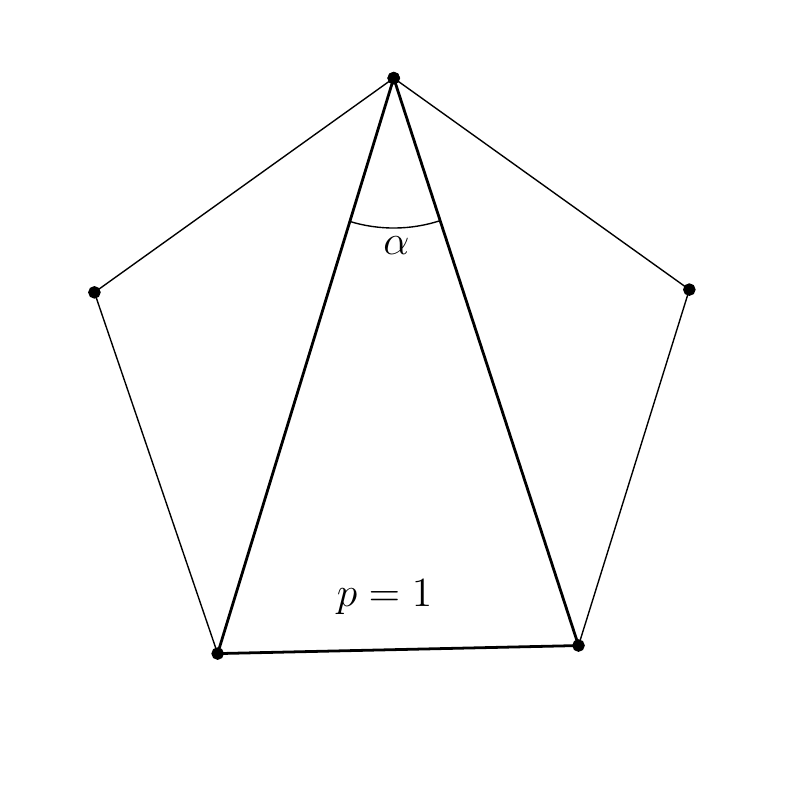}
\includegraphics[scale=.49]{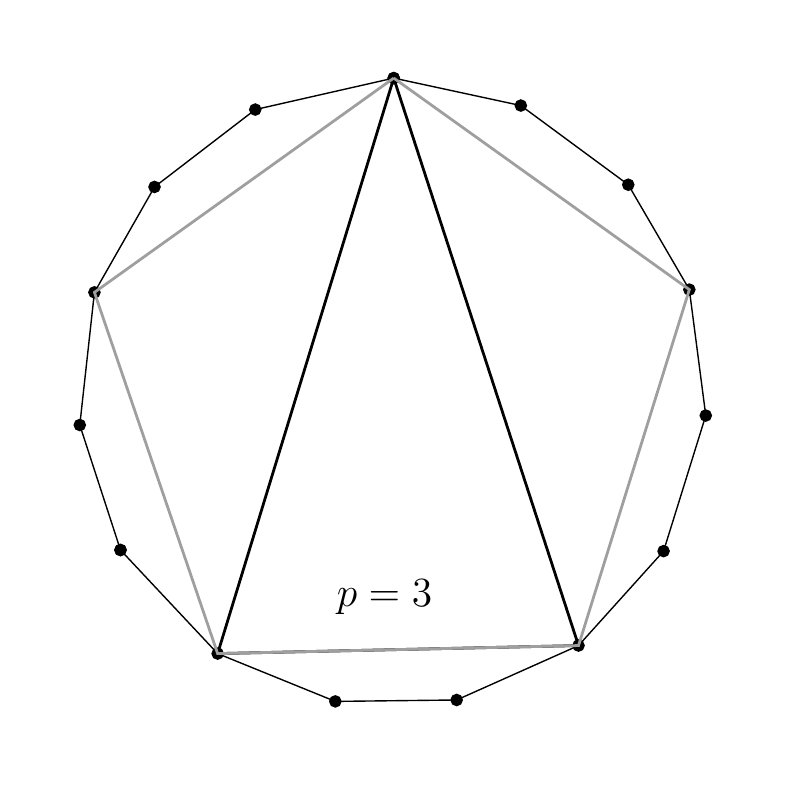}
\includegraphics[scale=.49]{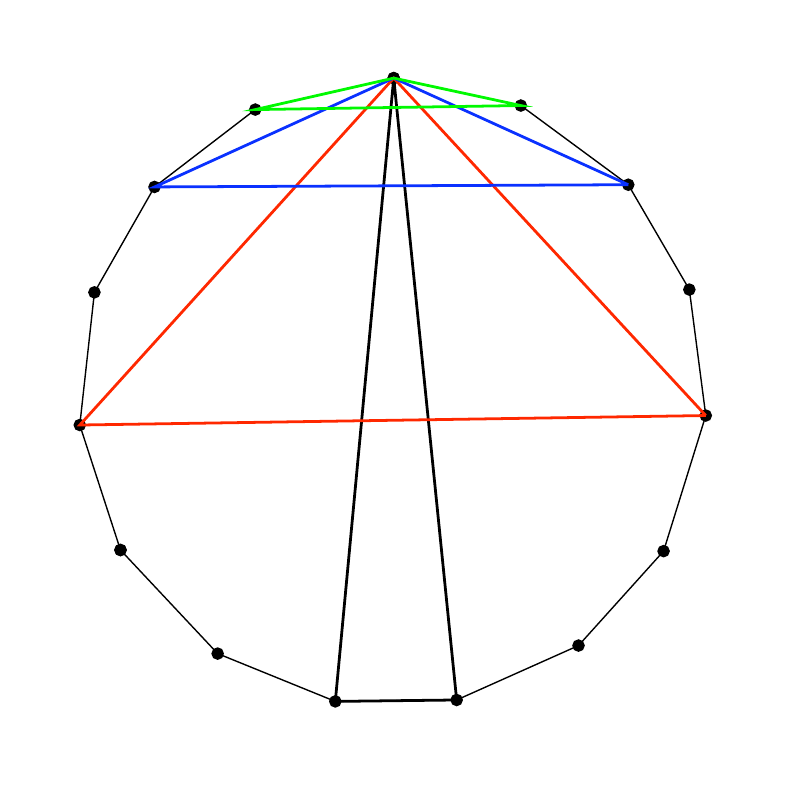}
\caption{Illustration of special triangles. The isosceles triangle with apex angle $\alpha$ is special, since it fits in the pentagon on the left, and satisfies the requirement of being relatively prime (the ratio is $2/5$).  
The same triangle does not satisfy this requirement in the 15-gon in the middle (ratio: $6/15$).   On the right
we illustrate the special triangles for the 15-gon (1,2,4,7 / 15).}
\label{fig:special}
\end{figure}

Any special triangle $T$ is particular type of isosceles triangle with vertices belonging to $S$.
What makes $T$ special is that its vertices belong to a (potentially incomplete) regular $k$-gon of $S$, and furthermore
the number of $k$-gon vertex positions skipped by the non-base sides of $T$ is relatively prime to $k$ (see
Figure~\ref{fig:special}).
In this section we give bounds on the probability that a triangle is special. We also
explain how a special triangle is uniquely associated to one regular $k$-gon.

\noindent Let $p(k\bot y)$ denote the probability that $k$ is relatively prime to $y$.
\begin{lemma}
\label{relativelyprimelemma} Let $k$ be an integer satisfying $3{\leq}k{\leq}n$, and let $y$ be
an  integer chosen uniformly at random in the interval 
$[1, \lceil\frac{k}{2}-1 \rceil ]$.  
  Then  $p(k\bot y) = \Omega(\frac{1}{\log n})$.
\end{lemma}
\begin{proof}
We know from the {\em prime number theorem} (Gauss, 1792;
see~\cite{havil})
that the number of primes smaller than
any given integer $x$ is $\Theta({x\over \log x})$. 
Thus there are  $\Theta({k \over \log k})$ primes in the
range $[\sqrt{k} ,\lceil \frac{k}{2}-1 \rceil )$.   Any such prime $y$  satisfies $k\bot y$.
The probability that some $y$ is chosen uniformly at random is 
$\Theta(\frac{1}{\log k}) = 
\Omega(\frac{1}{\log n})$.
\end{proof}
Lemma~\ref{relativelyprimelemma} tells us that in a regular $k$-gon, among all isosceles 
triangles with a given apex, the probability of randomly choosing a special triangle is
$\Omega(\frac{1}{\log k})$.
The following Lemma confirms that we can quickly associate a given special triangle to its
unique $k$-gon.
\begin{lemma}
\label{lem:unique}
If a regular $k$-gon has a special triangle $T$, then $T$ is not special for any other regular $k'$-gon.
\end{lemma}
\begin{proof}
Suppose that each non-apex side of $T$ skips $d$ vertices of the $k$-gon.
Let $\theta$ be the apex angle of $T$.  This angle can be viewed as a function of $\frac{d}{k}$.

If $T$ can be embedded on some other $k'$-gon, where $d'$ vertices are skipped, 
the ratio $\frac{d'}{k'}$ must also equal $\theta$.
Since $T$ is special for the $k$-gon, we know that $k\bot d$.  Therefore for any other
regular $k'$-gon, we have $\frac{k'}{d'}=\frac{ck}{cd}$, where $c>1$.  This means that
$T$ does not satisfy the property of being relatively prime for other regular polygons.
\end{proof}

\begin{lemma}
\label{lem:table}
Let $T$ be chosen uniformly at random
from the set of all isosceles triangles formed by points of a regular $k$-gon, where $k$ is
not given as input.
After $O(n^2)$ preprocessing, we can determine the value $k$ in $O(1)$ time
 with probability $\Omega(\frac{1}{\log n})$.
\end{lemma}
\begin{proof}
In $O(n^2)$ time, we can construct a table containing the angle of every special triangle for
every regular $k$-gon, for $k{\leq}n$.   This is done as follows.
For an isosceles triangle present in a $k$-gon, let the {\em isosceles ratio} be the side length of the triangle divided by $k$.
For every $k$-gon, we can construct a list of all isosceles ratios, in $O(k)$ time.
The total time is quadratic over all values of $k\leq n$.  Then, we eliminate items with the same ratio, except for the one
created by an irreducible fraction.\\

 By Lemma~\ref{lem:unique}, the elements of the list are unique.
Thus, if we are given a special triangle, we can use binary search (or hashing for $O(1)$-time) to locate
the $k$-gon for which the triangle is special.

Now suppose that each non-apex side of $T$ skips $d$ vertices of the $k$-gon.
Clearly $d$ is in the range $[1, \lceil \frac{k}{2}-1\rceil ]$.
By Lemma~\ref{relativelyprimelemma}, $p(k\bot d) = \Omega(\frac{1}{\log n})$.  
%
%
\end{proof}

\begin{lemma}
\label{big-gon-many-iso}
For any ${\geq} n^{\alpha}$-gon $G$, at least $1 \over n^2$ of the
isosceles triangles formed by vertices of  $S$ are in $G$.
\end{lemma}

\begin{proof}
According to~\cite{pach}, the number of isosceles triangles among $n$
points in the plane is $O(n^{2{+}2\alpha{+}\epsilon})$. Any ${\geq}
n^{\alpha}$-gon $G$ defines $\Theta(n^{2\alpha})$ isosceles triangles. Thus,
at least $1/n^2$ of the isosceles triangles from $S$ have all three points in $G$.
\end{proof}

\noindent Lemmas~\ref{relativelyprimelemma} and~\ref{big-gon-many-iso} directly imply the following.
\begin{cor}
\label{c4}
For any ${\geq} n^{\alpha}$-gon $G$, $1 \over n^2 \log n $ of the
isosceles triangles are special triangles of $S$ associated with $G$.
\end{cor}

\subsubsection{Using special triangles to test the existence of regular polygons}
\label{testing}
We have already established that a special triangle is associated to a unique $k$-gon, but 
this does not mean that the  $k$-gon actually exists in $S$.   In this section we show that the total cost
of such verifications is expected to be on the order of the total complexity of $k$-gons in a set.
In other words, false  verifications do not cost too much.

\begin{lemma}
\label{l2}
The sum of complexities of all ${\geq} n^{\alpha}$-gons is at most
$n^{2{+}\alpha+\epsilon}$.
\end{lemma}

\begin{proof}
Let $\kappa_i$ be the number of $i$-gons in a fixed set $S$ of $n$ points.
The sum of the complexities of all ${\geq} n^{\alpha}$-gons is
$\sum_{i=\lceil n^{\alpha}\rceil}^n i \kappa_i$. Any $k$-gon generates at most $O(k^2)$
isosceles triangles of which at most  $O(\frac{k^2}{\log n})$ are special.
Thus there are at most  $\sum_{i=\lceil n^{\alpha}\rceil}^n \frac{i^2 \kappa_i}{\log i}$
special triangles. We know from 
\cite{pach} that there are at most $O(n^{2{+}2\alpha{+}\epsilon})$ distinct isosceles triangles.
Since special triangles are distinct, 

\[ \sum_{i=\lceil n^{\alpha}\rceil}^n \frac{i^2 \kappa_i}{\log i} = O(n^{2{+}2\alpha{+}\epsilon}) \]

\noindent
which since $n^{\alpha} {\leq} i {\leq} n$ gives

\[ \frac{n^{\alpha}}{\log n}\sum_{i=\lceil n^{\alpha}\rceil}^n i  \kappa_i = O(n^{2{+}2\alpha{+}\epsilon}) \]

\noindent and dividing and absorbing the $\log$ into the $\epsilon$ gives

\[ \sum_{i=\lceil n^{\alpha}\rceil}^n {i \kappa_i} = O(n^{2{+}\alpha + \epsilon}) \]

\noindent This last equation is exactly the statement of the lemma.
\end{proof}

\begin{defini}
If a $k$-gon has at least $k/2$ vertices in $S$, then it is \emph{more-than-half-full}.
Otherwise it is \emph{less-than-half-full}.
\end{defini}

\begin{cor}
\label{c2}
The sum of complexities of all more-than-half-full ${\geq} n^{\alpha}$-gons
in a set $S$ of $n$ points is at most $n^{2{+}\alpha+\epsilon}$.
\end{cor}

\begin{proof}
This is an easy variant of Lemma~\ref{l2}, as the more-than-half-full
condition only affects constants in the asymptotic notation.
\end{proof}

\begin{lemma}
\label{l3}
Let $T$ be a special triangle, associated to a $k$-gon $P$.  We can decide if  $P$ is less-than-half-full in $S$ in $O(1)$ expected time.  Otherwise if it is more-than-half-full, we decide if it is completely full in $O(k)$-time.  
\end{lemma}
\begin{proof}
The center of $P$ is found from the circumcenter of $T$ in constant time.


Given $T$, the center of the polygon, and the value of $k$ determined by Lemma~\ref{lem:table},
we can compute 
the location of any vertex of $P$ in constant time.  We begin checking the $k{-}3$ unconfirmed vertices in random order, without replacement.  If any vertex
is not present we terminate the procedure. Otherwise we output the candidate $k$-gon.

Checking an entire $k$-gon takes $O(k)$ time, so if $P$ is more-than-half-full, our claim trivially
holds.  If $P$ is less-than-half-full,
over half of the $k{-}3$ tests will fail. Since the tests are ordered
randomly, the expected number of tests is at most $2$.  Thus we expect to
spend only $O(1)$ time in this case.
\end{proof}


\subsubsection{The algorithm}
\label{alltogether}
So far, we have shown that we can identify every large regular polygon in a point set, if
we manage to find a special triangle belonging to each such polygon.
We have hinted that these special triangles will be found by selecting ``enough" random isosceles triangles.
The following theorem addresses this issue and states our main result for large regular polygons.

\begin{theorem}
\label{big-gons}
With high probability, we can find all  ${\geq}n^\alpha$-gons in a set $S$ of $n$ points in the 
plane in expected time $O(n^{2{+}\alpha+\epsilon})$.
\end{theorem}

\begin{proof}
The {\em Coupon Collector problem} tells us that if we randomly select items out of a set of $n$ elements, with replacement, we should expect to have observed all $n$ items after $O(n\log n)$ selections.    This also means that if we have $n$ disjoint sets of elements, and the probability of selecting something from each set is
equal (i.e. $1/n$), then we expect to have obtained a sample from each set in $O(n\log n)$ time.
More generally,  if the minimum probability of sampling from one of the disjoint sets is
$1/p$, then we expect to have sampled from all sets in $O(p\log p)$ time.

By Corollary~\ref{c4},   $\frac{1}{n^2\log n}$ of all isosceles triangles in $S$ are
special, for each ${\geq}n^\alpha$-gon.
So we set this as $p=n^2\log n$ to conclude that we expect to have sampled a special triangle for each
such polygon in $O(n^2\log n\cdot\log(n^2\log n)) = O(n^2\log^2 n)$ time.

We select $\Theta(n^2 \log^2 n)$ random
isosceles triangles formed from the vertices of $S$. 
So, with  constant probability we will obtain at least one special triangle for every ${\geq}
n^{\alpha}$-gon. 
With quadratic pre-processing time,
we can select each random isosceles triangle in constant time (Lemma~\ref{lem:random}).

By Lemma~\ref{lem:table}, in constant time we can determine if one of our random isosceles triangles is
special, and if so we can determine the position of its unique candidate $k$-gon.
Lemma~\ref{l3} explains how we perform a verification for each special triangle (or, in other words,
for each potential $k$-gon).
We expect to spend $O(1)$ time  for each less-than-half-full $k$-gon, which means
$O(n^2 \log^2 n)$ time for all such cases. 
The cost of verification for more-than-half-full $k$-gons is proportional to their size.
We can avoid verifying the same $k$-gon twice by constructing a hash table, using the center
and top vertex coordinate as keys.
Corollary~\ref{c2} states that the total size of all
more-than-half-full ${\geq} n^{\alpha}$-gons is $O(n^{2{+}\alpha{+}\epsilon})$. 
\end{proof}





\section{Combinatorial questions}

A purely combinatorial question is the following: what is the maximum complexity of all of the regular polygons in a set of $n$ points in the plane.
Lemma~\ref{l2} bounds the complexity with respect to regular ${\geq}n^\alpha$-gons to be $O(n^{2+\alpha+\epsilon})$. Trivially, the complexity of all regular ${\leq}n^\gamma$-gons is $O(n^{2+\gamma})$, for any $\gamma$. Thus, the total complexity of all regular polygons is $O(n^{2+\alpha+\epsilon})$. This bound is highly dependent on the number of isosceles triangles, but perhaps a tighter bound is possible by using more properties of regular polygons, instead of simply treating them as generators of isosceles triangles.

We note that an improvement of the bound of the total complexity of regular polygons from $O(n^{2+\alpha+\epsilon})$ to $O(n^{2+\frac{2}{3}\alpha+\epsilon})$ would give a corresponding speedup in our algorithm (by increasing the number of random triangles and reducing the cutoff between small and large polygons). This is because the current bottleneck in our algorithm is the last step, the successful verification for special triangles.

Bra{\ss} observed that one can construct a set of points with $c_kn^2$ regular $k$-gons, for any constant $k$, which gives a lower bound $\Omega(n^2)$ for the number of regular polygons in a point set. An upper bound on the number of regular polygons is $O(n^2 \log n)$. Every pair of points defines at most two $k$-gons, thus there are at most $\frac{2n^2}{k}$ $k$-gons.  The total number of regular polygons is  therefore at most $\sum_{k=3}^n \frac{2n^2}{k} = O(n^2 \log n) $.

While the gap between the lower and upper bounds is much larger for the complexity of polygons rather than the number of regular polygons, it remains unknown if  these two quantities differ asymptotically.
\section{Acknowledgments}

We would like to thank Boris Aronov, Jean Chapelle and Erik Demaine for
interesting discussions about chords, polygons, and Euclid in general.


\bibliographystyle{abbrv}

\begin{thebibliography}{1}

\bibitem{brass}
Peter Bra{\ss}.
\newblock On finding maximum-cardinality symmetric subsets.
\newblock {\em Computational Geometry --- Theory and Applications}, 24(1):19--25,
  2003.

\bibitem{sweep}
Herbert Edelsbrunner and Leonidas Guibas.
\newblock Topologically sweeping an arrangement.
\newblock {\em J. Comput. Syst. Sci.}, 38(1):165--194, 1989.

\bibitem{havil}
Julian Havil.
\newblock {\em Gamma: Exploring Euler's Constant.}
\newblock Princeton, NJ: Princeton University Press, 2003.

\bibitem{pachagarwal}
J\'{a}nos Pach and Pankaj K. Agarwal.
\newblock {\em Combinatorial geometry}.
\newblock Wiley-Interscience, 1995.

\bibitem{pach}
J\'{a}nos Pach and G\'{a}bor Tardos.
\newblock Isosceles triangles determined by a planar point set.
\newblock {\em Graphs and Combinatorics}, 18:769--779, 2002.

\end{thebibliography}

\end{document}